# An Inverse Problem Study: Credit Risk Ratings as a Determinant of Corporate Governance and Capital Structure in Emerging Markets: Evidence from Chinese Listed Companies

ManYing Kang and Marcel Ausloos *

School of Business, University of Leicester, Leicester LE1 7RH, UK; kangmanying@foxmail.com (M.K.)
* Correspondence: ma683@le.ac.uk

**Abstract:** Credit risk rating is shown to be a relevant determinant in order to estimate good corporate governance and to self-optimize capital structure. The conclusion is argued from a study on a selected (and justified) sample of (182) companies listed on the Shanghai Stock Exchange (SHSE) and the Shenzhen Stock Exchange (SZSE) and which use the same Shanghai Brilliance Credit Rating & Investors Service Company (SBCR) assessment criteria, for their credit ratings, from 2010 to 2015. Practically, 3 debt ratios are examined in terms of 11 characteristic variables. Moreover, any relationship between credit rating and corporate governance can be thought to be an interesting finding. The relationship we find between credit rating and leverage is not as evident as that found by other researchers from different countries; it is significantly positively related to the outside director, firm size, tangible assets and firm age, and CEO and chairman office plurality. However, leverage is found to be negatively correlated with board size, profitability, growth opportunity, and non-debt tax shield. Credit rating is positively associated with leverage, but in a less significant way. CEO-Board chairship duality is insignificantly related to leverage. The non-debt tax shield is significantly correlated with leverage. The correlation coefficient between CEO duality and auditor is positive but weakly significant, but seems not consistent with expectations. Finally, profitability cause could be regarded as an interesting finding. Indeed, there is an inverse correlation between profitability and total debt (Notice that the result supports the pecking order theory). In conclusion, it appears that credit rating has less effect on the so listed large Chinese companies than in other countries. Nevertheless, the perspective of assessing credit risk rating by relevant agencies is indubitably a recommended time dependent leverage determinant.

## 1. Introduction

Ashbaugh-Skaife et al. (2006) investigated whether (894 S&P) firms with strong corporate governance benefit from higher credit ratings relative to firms with weaker governance. After controlling for firm-specific risk characteristics, it was found that "credit ratings are negatively associated with the number of shareholders and CEO power, and positively related to takeover defenses, accrual quality, earnings timeliness, board independence, board stock ownership, and board expertise". The data used was from the 2003 proxy season covering the board and committee structures of firms for the 2002 fiscal year, and pertain to a well-established (S&P) market.

On one hand, we can turn the problem upside-down: Is credit rating, in particular called credit risk rating, shown to be a valuable determinant for corporate governance and capital structure?

This connects to a puzzling question, also raised by Ashbaugh-Skaife et al. (2006), why some firms appear to be willing to bear additional debt financing costs by not practicing good governance. In fact, the sensitive question has been raised by Kisgen (2009): "Do firms target credit ratings or leverage levels?"

According to Kisgen (2006), the company adjusts capital structure on the basis of different credit rating level; elsewhere, Kisgen (2009) showed that a manager engages in capital structure behaviors such as to set a minimum credit rating level goal, so that the company is more likely to decrease its debt as the rating downgrades. Kisgen and Strahan (2010) implied that the companies' cost of debt capital is impacted by the ratings-based rules on bond investments. Furthermore, according to Bosch and Steffen (2011), when the company is not rated, no capital will be provided by non-bank investors, and loan shares will be increased.

Prior literature investigating firms' credit ratings and debt costs models the cost of debt as a function of issue characteristics and issuer risk attributes (see e.g., Horrigan 1966; Kaplan and Urwitz 1979) while ignoring the governance mechanisms that are put into place to safeguard the assets of the firm and ensure that bondholder interests are well served.

Sengupta (1998) and Bhojraj and Sengupta (2003) explored the effects of corporate governance on debt ratings and cost of debt financing, within a restricted set of governance variables. Sengupta (1998) finds a negative relationship between firms' disclosure quality ratings and the cost of debt financing as reflected in realized yields on new debt issues, while Bhojraj and Sengupta (2003) find that firms with a higher percentage of outside directors on the board and with greater institutional ownership get higher ratings on their new debt issues.

Thus, to the extent that governance is an important determinant of credit ratings, it can have a significant effect on firms' external financing costs. The matter is rather relevant for young markets, like China. Thus, we extend previous studies toward an "emerging market", and turn the question around.

On the other hand, without considering, *per se*, credit rating, it can be admitted that the factors of corporate governance including board size, auditor, outsider directors and duality do affect capital structure. In fact, Claessens et al. (2002), Saad (2010), San and Heng (2011), Haque et al. (2011), Ahmed Sheikh and Wang (2012), Morellec et al. (2012) and Ali et al. (2014) provided an empirical evidence that corporate governance influences capital structure both not only in developed but also in developing countries, even though Modigliani and Miller (1958) had argued that a firm capital structure is unrelated to a firm value in an efficient market.

More specifically, Wen et al. (2002), Chen (2004) and Ruan et al. (2011) have studied the relationship between corporate governance and capital structure of Chinese companies. However, such prior researches limited their study to the selected sample of China's civilian-run listed firms over the 2002–2007 time intervals. Other samples in these studies stopped data ca. 1996. Thus, one can say that they are all somewhat outdated.

Therefore, agreeing with Dasilas and Papasyriopoulos (2015), beyond such studies, the relationship between corporate governance and capital structure is still leaving space for exploration, in particular on the Chinese economy.

However, we insist that following our literature review several variables of interest appear to be missing and should be considered; several will be encountered below. One missing ingredient pertains to credit rating influence on the relationship between corporate governance and capital structure. The matter seems very relevant since credit rating would admittedly impact a manager's decision on growth and other performance matter strategies. In fact, several corporate finance textbooks have emphasized that credit rating may impact capital structure (Kronwald 2009; Agarwal 2013; Modina 2015).

Merton (1974) was the first to show that bankruptcy is starting at debt's maturation only when the solvency is deficient to satisfied present duty. Credit ratings are thus the important assessments for forecasting company debt default risks, when considering capital structure decision (Molina 2005). Interestingly, Graham and Harvey (2001) showed that credit rating ranks higher than any other variables in traditional capital structure models when issuing debt.

Several researches show that the determinant of credit rating is different in emerging and developed financial markets. The possible reason is that the rater considers that any financial ratio in a developing country is less authentic than in developed countries. For example, Ferri and Liu (2005) find that in developed markets, credit rating is negatively related to corporate rating, whereas the correlation is positive in developing markets. Ferri et al. (2001) discovered that banks and corporate rating in developing countries are significantly related *in fine* to the sovereign rating. Therefore, it goes without saying that the "developing countries" should adjust their criteria to different credit ratings schemes, though could be different from those in developed markets.

Specifically considering the China case, it has been pointed out that that the Chinese credit rating system present "problems". According to Firth et al. (2009), China banking systems lack unified and reliable interbank information, beside a valid national credit rating system. Zhen (2013) further agrees that China is short of reliable and independent credit rating agency, leading to problematic bond markets. However, these problems are mitigated and completed with China's increasing emphasis on credit. Yet, according to Poon and Chan (2008), China credit rating agency overoptimistically provides credit rating.

The present study emphasizes the relationship between credit rating, and capital structure plus corporate governance by using samples of "large listed companies" with A-shares on the Shenzhen Stock Exchange (SZSE), out of 2484, from 2010 to 2015. We claim that there is no similar research that uses the data examining corporate governance and credit rating to test capital structure, whence providing a novel view of capital structure determinant as done below. Moreover, the present study uses recent data from China, whence presents modern financial information. Finally, the benefit of using large listed firms allows one to more easily capture the fundamentals of equity and debt market. Thus, we extend previous studies not only toward an "emerging market", and turn the question around, but also suggest how to provide a novel view of capital structure determinant and optimize corporate governance—at least in such a market framework.

The paper is organized as follows. Section 2 imports some necessary background on the financial market in China for better focusing on specific constraints. Section 3 substantiates the newness of the present considerations through some literature review systematically going through usual determinant hypotheses**.** Section 4 summarizes the data source, together with its useful validity, and presents the methodology for arriving at the empirical results in Section 5, in which we assess how the relation between credit risk rating and regulation varies with the governance structure. Section 6 concludes and suggests.

**2. Background of the Financial Market in China**

*2.1. Overview of China's Capital Markets*

Shenzhen Stock Exchange was established on 3 July 1991 and Shanghai Stock Exchange was established on 19 December 1990. The Shenzhen Stock Exchange includes the main board, the SME board (initiated on 17 May 2004), and the Growth Enterprise Market (GEM) board (initiated on 30 October 2009). As calculated by the China Securities and Futures Statistical Yearbook 2012, at the end of 2012, there were 2484 firms listed on the main boards of the two Stock Exchange covering about US$ 3.8 trillion market capitalization.

General local stocks of China are "A-shares". The "B-shares" have about 107 listed companies and used other country currency, such as U.S. or Hong Kong, so that the market was mainly limited to the foreign investor. Domestic Chinese citizens could have B-shares from 2001; "qualified foreign institutional investors" could have A-shares from 2003. N.B. Total market capitalization of B-shares is less than 0.5% of total market capitalization on the Shenzhen Stock Exchange and Shanghai Stock Exchange.

*2.2. Corporate Governance Regulations*

The China Securities Regulatory Commission (CSRC) and the Chinese government are responsible for imposing corporate governance law. The CSRC and State Economic and Trade Commission published the "Code of Corporate Governance for Listed Companies in China" on 7 January 2002. The code includes 8 chapters. These Chapters describe the shareholder rights, the regulations for controlling shareholders, the regulations for the director and director board, the duty and responsibility of the supervisory board, the performance evaluation of the directors, of the supervisors, and of the management, and of the information disclosure.

Indeed, from 2003, each large-listed company is requested to have at least five directors and less than 19 directors. The law requires the company to have at least 1/3 independent directors on the board. The independent directors are not allowed to have a relationship with managers. Moreover, independent directors are not allowed to become one of the top ten shareholders nor own more than one percent of the firm stocks. Managers may be directors. If a listed company has a committee of compensation, audit, nomination, the independent directors should occupy more than 1/2 seats on the board. Each large-listed company elects the director and the supervisor by using the voting system. The regulation also requires that the board needs to have a meeting at least 2 times a year.

Chinese large-listed firms are requested to own at least 3 supervisors. The supervisors on the board must involve a broker of the shareholders. At least 1/3 of the supervisors have to be employees of the company; directors or senior managers are not allowed to become supervisors. The supervisors may attend the director board meeting as the conventioneers but a supervisor has no voting right. The supervisors have the right to suggest removing the director or manager if they break rules.

There is no restriction on the company's leadership structure. The company is free to choose the duality structure or non-duality structure, i.e., the former means that one person can be the CEO and the board chairman simultaneously.

**3. More General Literature Review and Determinant Hypotheses**

*3.1. Corporate Governance Factors*

The majority of studies about the correlation between capital structure and corporate governance concentrate on "internal factors" which include board size, outside directors, auditor and duality of CEO, for example.

3.1.1. Board Size

Pfeffer (1972) was the first who discovered that board size is related to leverage: Pfeffer (1972, 1973) and Provan (1980) demonstrated that board size was associated with a firm's ability to extract critical resources such as amount of budget, external funding and leverage from its "environment". Goodstein et al. (1994) examined the conflict between the institutional, governance, and strategic functions of boards. Following a test on how high levels of board size and diversity, traditionally associated with optimal institutional and governance performance of boards, affect the board's ability to initiate strategic changes during periods of environmental turbulence, the authors suggest that board diversity, whence size, may be a significant constraint on strategic change.

Kyereboah-Coleman et al. (2006), Abor (2007), Bopkin and Arko (2009) also find that board size is positively related to leverage (in Ghana) and indicate that a large board employs high leverages to raise a firm value. It is also argued that a large board meets problems for reaching agreements, implying poor corporate governance.

By contrast, other researchers discovered that debt ratio is negatively related to board size and claim that a large board size could force a manager to keep a low leverage policy in order to raise the company performance (Berger et al. 1997). In fact, Anderson et al. (2004 discovered that a board size is negatively related to the cost of debts. Guest (2009) looked at the impact of board size on firm performance in the UK. He found that board size has a strong negative impact on profitability, (and on Tobin's Q and share returns), but no evidence that firm characteristics that determine board size in the UK lead to a positive relation between board size firm and performance. In contrast, he found that the negative relation is stronger for large firms—which tend to have large boards. Since in China, as mentioned here above, part of the board is held by the shareholders, one expects that a large board be rarely occupied by the main shareholders.

Thus, it is interesting to verify whether board size is negatively or positively related to leverage.

### 3.1.2. Outsiders

By definition, an outsider on the firm's board of directors is not an employee or a stakeholder in the firm.

It has been pointed out that if one has many outsiders on the board, monitoring would be stricter (Weisbach 1988; Wen et al. 2002; Morellec et al. 2012). This is interpreted through considering (Jensen 1986) that such outsiders stay away from performance burden itself—since connected with responsibility.

Berger et al. (1997) and Wen et al. (2002) discovered that outsiders are negatively related to leverage and indicated that a manager pursues lower leverage when he/she faces strong corporate governance. Anderson et al. (2004) found that board independence is negatively related with the cost of debts. Morellec et al. (2012) also showed that there is a negative correlation between board independence and agency cost.

However, Pfeffer (1972) and Pfeffer and Salancik (1989) pointed out that an outsider is positively related to leverage, explaining that the outsider could increase debts and equities of a company by reducing the information asymmetry, or developing the companies' status and exploiting valuable resources. Some authors discovered that a company that owns higher leverage has more outsiders, whereas a company that has fewer outsiders in board experiences has lower leverages (Jensen 1986; Berger et al. 1997; Abor 2007). According to Bopkin and Arko (2009), board independence is positively correlated with debt ratio. However, the correlation is not significant. Kyereboah-Coleman and Biekpe (2006) also showed that there is a positive correlation between total leverage and outsiders, but found the correlation to be not significant.

Chen and Bradley (2015) had a mitigated set of findings about board composition effects: board independence decreases the cost of debt when credit conditions are strong or leverage is low, but it increases the cost of debt when credit conditions are poor or leverage is high; moreover, independent directors set corporate policies that increase firm risk.

It can thus be expected that the presence of outsiders be negatively related to leverage, because of the traditional composition of the board of Chinese firms.

### 3.1.3. CEO Duality

Duality means that the CEO also chairs the board. Duality is evidently offering more power (Boyd 1995). Pfeffer and Salancik (1989) pointed out that the managers who have better discretion could have more capacity to carry out the decision and have more power to overcome the laziness and damping modes of an organization. Fama and Jensen (1983) discovered that duality impacts the company's decision but the correlation is rather insignificant. Kyereboah-Coleman and Biekpe (2006) discovered that short-term leverage and total leverage are negatively associated with CEO duality, but long-term leverage is positively related to the CEO duality, although the correlation is not statistically significant. In contrast, Fosberg (2004) pointed out that a company that does not have a unitary leadership structure uses more debts than a company that has a unitary leadership structure. He discovered that a company that has no unitary leadership structure has high debts.

Some studies found that duality is positively related to leverage (Abor 2007). Bopkin and Arko (2009) also show that CEO duality is positively correlated with leverage but the correlation is not significant, explaining that a traditional CEO is more likely to operate a company through debt capital rather than stock issuance.

The relationship between duality and leverage will be examined.

### 3.1.4. Auditor

The auditor factor seems always ignored by previous research when studying the relation between corporate governance and capital structure. KPMG, Ernst & Young, Deloitte and PWC are called the Big-four auditors (De Franco et al. 2011). The Big-four auditors obey strict auditing regulations when working with firms. The high quality and the good reputation of the auditor can lower the possibility of alliance and dishonest activities between auditor and firms (Dasilas and Papasyriopoulos 2015). According to De Franco et al. (2011), people believe that the firms, which have Big-four auditoring, would have stronger internal control. In fact, Karjalainen (2011) shows that the Big-four auditor could decrease the cost of debt capital.

Chen et al. (2011) also discovered that a reliable auditor could reduce the cost of equity capital in non-state owned enterprises advantageously more than in state-owned enterprises—in China, explaining that since the audit report could constrain management decision, it could reduce information risks and could bring tangible benefits in view of decreasing the cost of equity capital.

Thus, one expects that the Big-four auditor will be positively related to debt also in our sample.

### 3.2. Company Factors

Previous research tested the company variables of capital structure primarily on the basis of the pecking order theory (Myers and Majluf 1984) and the trade-off theory (Robichek 1967). Myers and Majluf (1984) advanced the pecking order theory, which assumes that the cost of debt rises with asymmetric information. The trade-off theory (Robichek 1967) supports that business connects with a hierarchy of financing resources and has a preference for internal financing. If a firm needs external financing, the firm would more likely choose debts than equities. Myers and Majluf (1984) also pointed out that increasing capital is a signal that the firm does not want to choose equities, because when the manager issues new equities (and since the manager is supposed to know more about the company than an investor), an investor would tend to believe that the manager will prefer to consult people who will overvalue the company (since in fact the manager would finally receive benefits from this overvaluation). The value relevance of financial reporting might not be always appreciated (Ali and Hwang 1999)—the more so in not too developed markets (Jianu et al. 2014).

### 3.2.1. Firm Size

Both the pecking order theory and the trade-off theory regard the firm size as the variable of capital structure. The firm size is an inverse proxy of the bankruptcy cost and the earning volatility (Rajan and Zingales 1995; cited in Dasilas and Papasyriopoulos 2015).

According to Fama and French (2002), the trade-off theory forecasts that a firm size is positively correlated with leverage since a large company is more diversified and has less volatile earning. Thus, a bankruptcy cost could be reduced by a less volatile earning, since the company could get more debts (Degryse et al. 2012). According to Palacín-Sánchez et al. (2013), the pecking order theory also forecasts that the firm size is positively correlated with leverage since a large company has high quality and reliable information, whence allowing a decreasing cost of debts. Other researches also discovered that the firm size is positively related to leverage (Chowdhury and Chowdhury 2010; Céspedes et al. 2010; Ahmed Sheikh and Wang 2012).

Thus, it is expected that it will be found that the firm size is positively related to leverage.

3.2.2. Asset Tangibility

A tangible asset is always regarded as a loan's guarantee, whence suggesting that there is a positive correlation between the asset tangibility and debts. Moreover, asset tangibility could reduce the bankruptcy cost and credit risks (Dasilas and Papasyriopoulos 2015). Thus, both pecking order theory and trade-off theory predict that asset tangibility is positively correlated with leverage; asset tangibility also could mitigate information asymmetry troubles (Degryse et al. 2012). Céspedes et al. (2010) and Korteweg (2010) obtained that a tangible asset is positively correlated with leverage.

Mateev et al. (2013) pointed out that the influence between tangible assets and either short-term or long-term debts are different. The authors found that there is a positive correlation between long-term debts and tangible asset, but a negative correlation between short-term debts and tangible asset. In fact, according to Hoff (2012), there is a positive correlation between long-term debts and long-term assets, but a negative correlation between short-term debts and long-term assets.

Thus, it is expected that there is a positive relationship between long-term debts and tangible asset, and a negative correlation between short-term debts and tangible asset. It is also expected that tangible assets are positively related to leverage in our sample.

3.2.3. Profitability

The evidence in the relationship between profitability and debts turns out to be different in different studies. The trade-off theory points out that profitability is positively correlated with debts, in order to decrease tax liabilities, whereas the pecking order theory argues that profitability is negatively correlated with debts (Mac an Bhaird 2010). According to Degryse et al. (2012), high profitability means financing could be through cash flows of the firm, so it reduces the possibility to get debts. Several studies prove that profitability is negatively related to debts (Ching et al. 2011; Sharma and Kumar 2011; Shubita and Alsawalhah 2012; Akoto et al. 2013; Mateev et al. 2013).

On the other hand, Abor (2005) proved that due to the large percentage of short-term financing, profitability is positively related to debts because profitable company regards debts as the primary financing source (Khan 2012). Gill et al. (2011) discovered that both long-term debts and short-term debts are positively correlated with profitability. Debt is positively correlated with the Islamic bank's profitability in Pakistan (Akhtar et al. 2011). According to Li (2003), profitability is positively correlated with the debt capacity of a company because debt ratio can raise the rate of returns. Zhou (2009) proves that both short-term debts and long-term debts are positively correlated with corporate performance in China.

Thus, it seems interesting to discuss whether profitability is positively or negatively related to debts.

3.2.4. Growth Opportunity

Myers (1977) proves that company borrowing is negatively related to the growth opportunity—because the outcomes come from the growth opportunity. According to the trade-off theory, due to the uncollateralized of growth opportunity, companies with growth opportunity have a tendency to borrow less than companies which have high tangibility. Therefore, the growth opportunity is negatively related to debt (Ahmed Sheikh and Wang 2011). Others, Deesomsak et al. (2004), Huang (2006) and Bae et al. (2011) also provide some evidence that the growth opportunity is negatively correlated with leverage.

However, the pecking order theory assumes that there is a positive connection between debt and growth opportunity. According to De Jong (1999), companies that have much growth opportunity prefer to increase their (new) resources. According to Mateev et al. (2013), the growth of companies would put pressure on internal resources and rather tend to find external resources. Other empirical studies by Chen (2004), Giannetti (2003) and Degryse et al. (2012) prove that the growth opportunity is positively related to debt.

Huafang and Jianguo (2007) discovered that the larger companies in China have greater disclosure, while the companies that have growth opportunity are reluctant to disclose information. The listed companies increase disclosure activities since they want to decrease the information asymmetry (Lang and Lundholm 2000). According to Easley and O'Hara (2004), reducing the information asymmetry could decrease the equity costs that motivate companies to invest by adopting the equities, this could decrease debts. Furthermore, the director who is also the primary shareholder of the company does not want to lose control of the company, whence such a director is more likely to look for internal resources (Dasilas and Papasyriopoulos 2015).

It is obvious that whether the growth opportunity is negatively or not related to debts should be discussed.

3.2.5. Non-Debt Tax Shields (NDTS)

DeAngelo and Masulis (1980) discovered that, in addition to the interest expenditure, the accounting depreciation and investment tax credits (NDTS) would give tax profits to the company. The authors pointed out that a company which has large NDTS could adopt a policy for reducing its debt amount. Wald (1999), from the ratio of depreciation expenditure to total asset, discovered that debt is negatively related to the NDTS. Sogorb-Mira (2005) and Kolay et al. (2011) found that leverage is negatively correlated with the non-debt tax shields and in favor of the substitutability assumption of DeAngelo and Masulis (1980), the NDTS could replace the tax shield profits of leverage.

On the other hand, Ali (2011), Barakat and Rao (2012) and Anandasayanan et al. (2013) discovered that debt is positively correlated with the non-debt tax shields. In contrast, Degryse et al. (2012) discovered that NDTS is negatively correlated with long-term debt, but is positively correlated with short-term debt.

The sign of the correlation between debt and non-debt tax shields should thus be examined.

3.2.6. Firm Age

Company age means the number of years that a company has been in operation.

The pecking order theory argues that age is negatively correlated with debts because a longer age means that a company could create more internal finances, whence decreasing the requirement for external resources. According to Mac an Bhaird (2010), the external equities are negatively related to age of companies; his results also imply that aged companies take better advantage of short-term debts than younger companies. Ahmed et al. (2010) provide evidence that age of firms is negatively correlated with debt ratio in Pakistan. Noulas and Genimakis (2011) also provide evidence that there is a negative correlation between the measure of leverage and age of companies in Chinese listed companies. Palacín-Sánchez et al. (2013) pointed out that a young company could have to adopt debts when faced with restrictions to increase its finance, as retained in its first year.

In our case, it is thus expected that age of a company be negatively related to leverage.

*3.3. Credit Rating Factors*

The studies comparing the relationship between credit rating and capital structure are not fully researched (Dasilas and Papasyriopoulos 2015). According to Graham and Harvey (2001), the great credit rating and asset flexible finance are the most significant variables which influence debt policies where the former is the second most concerning factor for the company. Kisgen (2006) extended the research of Graham and Harvey (2001). He hypothesized the capital structure-credit ratings assumption. He assumed that credit rating is a material factor which should be considered by directors of a company when deciding the capital structure of the company; he argued that different profits are related to different credit rating levels. He found that credit rating impacted the capital structure determination on the United States markets (from 1986 to 2001). He also discovered that the firms which are close to a credit rating upgrade or downgrade issue less debt relative to equity than firms not near a change in rating. Kisgen (2009) examined the influence of recognized alteration in the company's credit ratings on capital structure, not just examining the outcome of the company being assessed to obtain a downgrade or upgrade.

However, Kemper and Rao (2013) argued that credit rating is not the first consideration in a capital structure determination, even though the company (its CFO) also needs to pay attention to the importance of credit rating in deciding the capital structure policy. In contrast, Drobetz and Heller (2014) discovered that the ability to meet financial obligations by publicly listed firms in Germany receives the most attention by rating agencies in assessing the creditworthiness of companies. Furthermore, "the level of debt held by firms is the second most important factor in deriving a credit rating". Profitability does not significantly affect the rating assessment. Interestingly, Drobetz and Heller also observed that the German company credit rating is "to some extent" predominated by soft facts, thus by quite other "measures" that financial risk and business risk factors.

Thus, our present study regards credit rating as an extra factor which could influence capital structure. We focus on testing the relationship between capital structure and credit rating as well as corporate governance. However, this paper does not examine or discuss how credit rating affects corporate governance, but it seems valuable to pay attention to the relationship between credit rating and corporate governance.

It is assumed that a company that has low creditworthiness would face problems in the debt market and undergo higher debt costs. However, a company which has a high creditworthiness would more easily get some (bank) credit and thereby could lower its cost of debts.

Therefore, it seems that one may expect a positive relationship between credit ratings and leverage.

## 4. Data and Methodology

### *4.1. Dataset*

The samples of this study are the large listed companies which come from Shanghai Stock Exchange (SHSE) and Shenzhen Stock Exchange (SZSE) from 2010 to 2015. However, different firms use different credit agencies and different credit rating systems. Thus, for coherence purposes, in order to analyze the data under the same credit rating system, this study is restricted to a sample for which credit ratings use the assessment by the Shanghai Brilliance Credit Rating & Investors Service Company (SBCR) in A-shares. Specifically, SBCR was established in July 1992 and is a credit rating agency with high qualifications and reputation, in China. In so doing, a total of 182 firms, evaluated by the SBCR in A-shares during 2010 to 2015, are selected. The raw data for variables describing capital structure and credit rating is collected from CSMAR, which is an authority certified database in China; the data of corporate governance is retrieved from SZSE. N.B. This study cannot include SME in the sampling because of incomplete credit rating data information on such SME, in China.

### *4.2. Variables*

#### 4.2.1. Capital Structure

This study follows the example of earlier researches, such as Degryse et al. (2012), Palacín-Sánchez et al. (2013) and Dasilas and Papasyriopoulos (2015). We also use three proxies to represent capital structure: debt ratio (DR), long-term ratio (LDR) and short-term ratio (SDR). The DR is the ratio of liability to total asset; LDR is the quotient between long-term debt and total asset; SDR is the quotient between short-term ratio and total assets (Degryse et al. 2012; Palacín-Sánchez et al. 2013; Mateev et al. 2013; Dasilas and Papasyriopoulos 2015). N.B. Debt ratio includes short-term ratio and long-term ratio, but not necessarily in a linear superposition way.

4.2.2. Corporate Governance

Corporate governance factors include (i) board size (BOARD), which is the logarithm of the amount of board officers, (ii) the number of outside directors (OUTSIDERS), i.e., those firm's board of directors, but neither employees nor stakeholders in the firm (Wen et al. 2002; Abor 2007). Other dummy variables are (iii) the CEO duality (DUALITY) if it exists, and (iv) the auditor (AUDITOR). The former variable uses 1 if the CEO is also the chairman in the board and 0 otherwise (Abor 2007). The latter variable uses 0 if the auditors of the firm include at least one member from the Big-Four accounting firms and 1 otherwise.

4.2.3. Firm-Specific Determinants of Capital Structure

The firm-specific determinants of capital structure include (v) the firm size (SIZE) which is the natural logarithm of total sales income (in CNY millions) here (Kemper and Rao 2013; Dasilas and Papasyriopoulos 2015). The net fixed assets divided by total assets of the company at final financial year is (vi) the asset tangibility (TANGIBILITY) (Palacín-Sánchez et al. 2013; Mateev et al. 2013; Dasilas and Papasyriopoulos 2015). The earning before interests and taxes divided by total assets is (vii) profitability (PROFIT) (Palacín-Sánchez et al. 2013; Mateev et al. 2013; Dasilas and Papasyriopoulos 2015). The market values of equities divided by the book values of total assets is (viii) the opportunities (GROWTH) (Kemper and Rao 2013; Dasilas and Papasyriopoulos 2015). The annual depreciation charge divided by total assets is taken as (ix) the non-debt tax shield (NDTS) (Degryse et al. 2012; Dasilas and Papasyriopoulos 2015). Finally, (x) the natural logarithm of the amount of years of firm operation is firm age (LOGAGE) (Palacín-Sánchez et al. 2013; Dasilas and Papasyriopoulos 2015).

4.2.4. The Credit Rating System

In May 2003, China Insurance Regulatory Commission, through the "Interim Measures" for the insurance company investment management of corporate bonds, recognized five credit rating firms: the Far East Credit Rating, CCXI, China Lianhe Credit Rating Company, Shanghai Brilliance Credit Rating & Investors Service Company (SBCR) and Dagong Global Credit Rating Company. We took SBCR as a qualified credit rating company, in order to filter different creditworthiness appreciation. The semi-qualitative semi-quantitative criteria can be found in SBCR Publications. Such criteria lead to three levels of risk.

In brief, the lowest level of risk is "AAA"; the highest default risk is "CCC". Based on such a gradation, the eleventh (xi) credit rating (CR) explanatory variable in this study uses a 4 quartile scale: a firm gets 4 points if found to be AAA, 3 points if AA+, AA or AA−, 2 points for A+, A or A−, 1 point for BBB+, BBB, BBB−. There is no company rated C or below.

*4.3. Model Specification*

This study uses the panel data analysis statistical method (Maddala, 2001). Practical information: the statistical analysis uses the STATA on computer Windows 10. Indeed, the analysis includes several cross-sectional variables and extends over several (six) years. Such a panel data analysis is adopted by a majority of previous researches because it is claimed to provide a greater depth analysis than other cross-sectional data analysis (Daskalakis and Psillaki 2008; Psillaki and Daskalakis 2009; Dasilas and Papasyriopoulos 2015; Dedu et al. 2014; Jaba et al. 2017); nevertheless none should be *a priori* rejected, due to their complementarity (García-Gallego and Mures-Quintana 2016). Firstly, according to Wooldridge (2010) and Hsiao (2014), this data analysis is rarely influenced by the multicollinearity among the representative data, whence can provide greater evaluations. Secondly, according to Arellano (2003) and Baltagi (2008), this model controls the existence of firm-specific outcomes in a regression analysis. Thirdly, according to Arellano (2003), Baltagi (2008) and Hsiao (2014), this model is superior in order to recognize and assess the outcomes which could not be only loosely examined in cross-sectional analysis or time-series analysis.

Thus, as in Degryse et al. (2012), Vermoesen et al. (2013), Mateev et al. (2013) and Dasilas and Papasyriopoulos (2015), the study adopts the fixed effects panel data analysis (Maddala 2001) because it could control the time-invariant variance of the variables (Baltagi 2008; Wooldridge 2010), taking into account O'brien (2007) warning. As such, the ordinary least squares (OLS) will be used to evaluate the coefficients in

$$\text{DEBT}_{i,t} = \beta_0 + \beta_1 \text{BOARD}_{i,t} + \beta_2 \text{OUTSIDERS}_{i,t} + \beta_3 \text{DUALITY}_{i,t} + \beta_4 \text{AUDITOR}_{i,t} + \beta_5 \text{SIZE}_{i,t} + \beta_6 \text{TANGIBILITY}_{i,t} + \beta_7 \text{PROFIT}_{i,t} + \beta_8 \text{GROWTH}_{i,t} + \beta_9 \text{NDTS}_{i,t} + \beta_{10} \text{LOGAGE}_{i,t} + \beta_{11} \text{CR}_{i,t} + \beta_{12} \text{ROE}_{i,t} + \varepsilon_{i,t} \quad (1)$$

The index i on DEBT pertains to the sort of studied variables: DR (Debt ratio, i = 1); LDR (Long-term debt ratio, i = 2) and SDR (Short-term debt ratio, i = 3), respectively, at time *t*. The ROE term origin, written in the above model, will be explained below in Section 5.2; at this stage, let $\beta_{12}$ = 0.

Moreover, Sun et al. (2002) proved that the partial government ownership is positively related to state owned enterprises (SOEs) performance. One could be rightly concerned by the fact that, in China, firms have political relationships with local government. Therefore, the availability of bank loan might depend on firms' ownership type, being SOE or not. This would introduce a $\beta_{13}$ term. For completeness, let us mention that we have considered such a possibility, but have found no significant result, whence have not explicitly introduced the variable in the above modeling, and have neglected the term thereafter.

## 5. Empirical Results

### 5.1. Capital Structure Determinants of Large-Listed Firms

Table 1 displays the descriptive statistics for capital structure variables, company variables, corporate governance factors and credit rating score. Recall that 182 companies are usually examined over 6 years, leading to *N* = 1092 data points. A few points are missing, due to a lack of corresponding raw data. From Table 1, one can see that the mean of total debt ratio, long-term debt ratio and short-term debt ratio distribution is 52%, 10% and 38%, respectively. The (integer) mean of the number of board members is 9; the percentage of outside directors is 37%, i.e., less than half of the board members are made of independent actors. The proportion of CEO duality occurs about 81%, i.e., in most companies the same person holds the CEO and board chair at the same time. About 9% of the companies hire Big-four auditors. The mean annual sales are about 789 millions (CNY). The tangibility, profitability, and non-debt tax shields mean is 22%, 6% and 2%, respectively. The mean number of operation time is about 32 months. The credit score mean is about equal to 3.

**Table 1.** Descriptive statistics of the dependent and independent variables.

| Variable | Definition | N | Mean | Median | SD | Min | Max |
|---|---|---|---|---|---|---|---|
| DR | Total debt/total asset | 1092 | 0.520 | 0.521 | 0.173 | 0 | 1.056 |
| LDR | Long term debt/total asset | 1089 | 0.102 | 0.069 | 0.113 | −0.006 | 0.581 |
| SDR | Short-term debt/total asset | 1089 | 0.377 | 0.381 | 0.174 | 0 | 1.004 |
| BOARD | Number of board members | 1006 | 9.147 | 9 | 1.926 | 0 | 18 |
| OUTSIDERS | Percentage of outside directors | 1004 | 0.367 | 0.333 | 0.053 | 0.250 | 0.667 |
| DUALITY | CEO and chairman office plurality | 1002 | 0.808 | 1 | 0.394 | 0 | 1 |
| AUDITOR | Big-four auditor | 1075 | 0.094 | 0 | 0.292 | 0 | 1 |
| SIZE | Annual sales (in millions CNY) | 1089 | 7.894 | 7.858 | 1.335 | 2.942 | 11.95 |
| TANGIBILITY | Tangible fixed asset/total asset | 1089 | 0.218 | 0.175 | 0.188 | 0.000 | 0.971 |
| PROFIT | Earnings before interests and taxes/total asset | 1089 | 0.0587 | 0.055 | 0.052 | −0.266 | 0.340 |
| GROWTH | Market value of equity/book value of asset | 1089 | 1.540 | 1.128 | 1.702 | 0 | 25.28 |
| NDTS | Annual depreciation/total asset | 1089 | 0.018 | 0.016 | 0.014 | 0 | 0.075 |
| AGE | Number of years of operation | 1092 | 16.14 | 16 | 5.53 | 0 | 37 |
| LOGAGE | Log of the Number of years of operation | 1091 | 2.781 | 2.773 | 0.385 | 0 | 3.611 |
| CR | Credit score | 1092 | 2.923 | 3 | 0.607 | 1 | 4 |
| ROE | Retained profits/equities | 1092 | 9.692 | 8.795 | 11.24 | −99.22 | 76.10 |

Table 2 shows the Pearson correlation coefficient between these factors. It can be observed that both long-term debts and short-term debts are positively correlated with total debt, but short-term debt is negatively correlated to long-term debt.

**Table 2.** Pearson correlation coefficient matrix between variables employed in the model regression analysis; the *, ** and *** represent statistical significance at the 0.1, 0.05 and 0.01 level, respectively.

| | DR | LDR | SDR | BOARD | OUTSIDERS | DUALITY | AUDITOR |
|---|---|---|---|---|---|---|---|
| LDR | 0.334 *** | | | | | | |
| SDR | 0.469 *** | −0.127 *** | | | | | |
| BOARD | 0.210 *** | −0.082 *** | −0.160 *** | | | | |
| OUTSIDERS | 0.077** | 0.143 *** | 0.034 | −0.311 *** | | | |
| DUALITY | 0.138 *** | 0.054 * | 0.019 | 0.114 *** | −0.091 *** | | |
| AUDITOR | 0.143 *** | 0.026 | −0.081 *** | 0.252 *** | 0.020 | 0.087 *** | |
| SIZE | 0.436 *** | 0.008 | 0.199 *** | 0.292 *** | −0.050 | 0.125 *** | 0.267 *** |
| TANGIBILITY | 0.009 | 0.331 *** | 0.031 | −0.008 | −0.021 | 0.028 | −0.084 *** |
| PROFIT | −0.306 *** | −0.105 *** | 0.002 | −0.135 *** | 0.033 | −0.095 *** | 0.003 |
| GROWTH | −0.420 *** | −0.216 *** | −0.130 *** | −0.144 *** | −0.023 | −0.151 *** | −0.108 *** |
| NDTS | −0.065 ** | 0.102 *** | 0.122 *** | −0.020 | −0.032 | −0.037 | −0.095 *** |
| LOGAGE | 0.163 *** | 0.062 ** | 0.007 | 0.010 | 0.113 *** | 0.050 | 0.010 |
| CR | 0.228 *** | 0.075 ** | −0.132 *** | 0.272 *** | −0.025 | 0.164 *** | 0.270 *** |
| | SIZE | TANGIBLITY | PROFIT | AGE | GROW | NDTS | CR |
| TANGIBILITY | 0.096 *** | | | | | | |
| PROFIT | −0.085 *** | −0.008 | | | | | |
| GROWTH | −0.314 *** | −0.161 *** | 0.123 *** | 0.013 | | | |
| NDTS | 0.155 *** | 0.782 *** | 0.002 | −0.058 * | −0.123 *** | | |
| LOGAGE | 0.192 *** | −0.123 *** | −0.186 *** | | | | |
| CR | 0.504 *** | −0.076 ** | −0.025 | 0.174 *** | −0.170 *** | −0.070 ** | |

As for corporate governance factors, both long-term debts and short-term debts are negatively correlated with board size. This indicates that the larger the board size, the less long-term debt and the less short-term debt would be, but there is more total debt. The result is consistent with above hypotheses and prior studies. Credit rating has a positive relationship with board size, CEO duality and auditor but a negative relationship with outsiders. Both Ashbaugh-Skaife et al. (2006) and Bradley et al. (2008) discovered that credit rating is positively related to board size, outsiders and auditor but negative related to CEO duality in the United States. Bhojraj and Sengupta (2003) found that outsiders have a positive relationship with credit ratings. Liu and Jiraporn (2010) observed a lowering of credit ratings when the CEOs have more decision-making power. Our results on the relationship between credit rating with outsiders and CEO duality are inconsistent with these researches. However, notice that Becker and Milbourn (2008) proved that credit rating coincides with good industry performance: a better performance leads to higher credit rating. Peng et al. (2007) supported the stewardship theory which represents that CEO duality is good for performance due to the unity of command it presents. Therefore, the CEO duality has a positive relationship with the performance which itself improves credit rating.

Institutional theory suggests that changes in organization, like, for example designating more outside directors, will result in a process that makes firms more similar rather than more efficient (DiMaggio and Powell 2000). In this line of thought, Peng (2004) suggests that outsider directors have a different influence on firm performance in Chinese firms, if the performance is measured by sales growth; moreover, he finds that outsiders have little impact on financial performance measured as the return on equity (ROE). Thus, a greater number of outsiders will lower the performance, itself being positively related to credit rating, thus will be lowering credit rating!

There is also a significantly positive correlation between the percentage of outside directors and total debt and long-term debt, indicating that the more outside directors on the board has a company, the more so is total debt and long-term debt. Notice that Guest (2008) proved that there is a positive relationship between the outsiders proportion (and board size) and debt ratio, justifying the finding through the argument that the supervisory and the advisory persons are the same outside director. It can be understood as follows: China's Company Law (2005) regulated that Chinese firms should have supervisory boards and management boards, both of them having a "social responsibility" in conducting financial operations. The monitoring functions impose that the management has the pursuit of the shareholder profits. Outsiders want to show their capability to the employer and have more independence than the insiders (Fama and Jensen 1983; cited in Guest 2008), whence a positive correlation factor.

The CEO duality is positively correlated to total debt and long-term debt, but weakly correlated to short-term debt, implying that the person who holds the CEO and board chairship, at the same time, usually prefers to choose long-term debt. Whether this is a merely psychological or financial corporate governance attitude is not further discussed here.

However, there is a positive correlation between the Big-four auditor choice and total debt, and interestingly, a negative correlation between the Big-four auditor and short-term debt. This means that the "high quality auditor" would rather influence an increase of long-term debt and a decrease of short-term debt. Notice that the correlation coefficient between CEO duality and auditor is positive but weakly significant (see below).

Consider the firm factors. As expected, the company size is positively correlated with total debt and short-term debt, whence it can be deduced that larger firms prefer short-term debt. The tangibility is positively but weakly correlated with long-term debt. It means that tangible assets may be regarded as the collateral to long-term debt. Profitability is negatively related to total debt and long-term debt, implying that if a company makes more profit, it would decrease its long-term debt. As expected, a negative correlation occurs between growth opportunities and debt. The non-debt tax shields are positively related to debt ratio, but negatively correlated to short-term debt and long-term debt ratios.

The firm age is also positively correlated with total debt and long-term debt—indicating that if a company operates has a longer operation time, the strategy is to prefer a long-term debt. This result contradicts above hypothesis. However, this seems to be explainable through an observation by the International Monetary Fund (Prasad and Ghosh 2005) which comments that the reason might be found in the fact that "small and young firms in emerging markets are likely to find debt cheaper than equity, since they may have easy access to credit" (Huissan and Hermes 1997; cited in International Monetary Fund, Prasad and Ghosh 2005).

The $β_i$ regression coefficients for the various debts, from 2010 to 2015, are found in Table 3.

**Table 3.** Regression outputs from 2010 to 2015.

| Variable | Expected Sign | DR | LDR | SDR |
|---|---|---|---|---|
| BOARD | − | 0.01 ** | −0.01 *** | −0.02 *** |
|  |  | (1.99) | (−3.39) | (−7.46) |
| OUTSIDERS | − | 0.35 *** | 0.20 *** | −0.05 |
|  |  | (4.15) | (3.25) | (−0.52) |
| DUALITY | ± | 0.01 | −0.00 | 0.01 |
|  |  | (0.58) | (−0.40) | (0.98) |
| AUDITOR | + | −0.02 | 0.02 | −0.04 ** |
|  |  | (−1.00) | (1.46) | (−2.30) |
| SIZE | + | 0.05 *** | −0.01 *** | 0.05 *** |
|  |  | (12.33) | (−3.74) | (11.39) |
| TANGIBILITY | + | 0.16 *** | 0.38 *** | −0.16 *** |
|  |  | (4.18) | (14.11) | (−3.84) |
| PROFIT | + | −1.02 *** | −0.14 ** | −0.37 *** |
|  |  | (−10.54) | (−2.06) | (−3.37) |
| GROWTH | − | −0.02 *** | −0.02 *** | −0.01 * |
|  |  | (−7.83) | (−7.29) | (−1.72) |
| NDTS | ± | −3.75 *** | −3.44 *** | 1.57 *** |
|  |  | (−7.49) | (−9.52) | (2.73) |
| LOGAGE | − | 0.04 *** | 0.02 ** | 0.00 |
|  |  | (3.04) | (2.08) | (0.32) |
| CR | + | 0.01 * | 0.02 *** | −0.07 *** |
|  |  | (−0.88) | (3.78) | (−6.91) |
| $R^2$ |  | 0.415 | 0.273 | 0.201 |
| F Test | F(16,982) = | 43.61 | 23.05 | 15.45 |
| Prob > F |  | 0.000 | 0.000 | 0.000 |

Prob > F represents the significance of the Joint test, all such test results are 0.000 < 0.01, illustrating that the significance results are over 1% level. T test is in parentheses. The *, ** and *** represent statistical significance at the 0.1, 0.05 and 0.01 level, respectively.

As for corporate governance factors, board size is negatively linked with long-term debts and short-term debts; the reason might be that a large board restricts the company to pursue the higher debts because higher debts could lower the company performance. This result is consistent with Anderson et al. (2004) research that larger board adopts less short-term and long-term debt.

However, board size is positively related to total debt. According to Bragg (2002), debt consists of short-term debt and long-term debt, but it also could be regarded as the current liabilities when debts are supposed to be paid off within more than 1 year or through some operating cycle. Current liability includes accounting payable, customer deposits, accrued tax, accrued wage and vacation pay and other accruals connect with current operations. Therefore, debt not only includes short-term and long-term debt, but also consists of other liabilities. Moreover, according to Abor (2007), larger board sizes would have more non-executive directors who might bring better management decisions and assist the company to get more resource—since (or because) these external board people might have "greater knowledge" and "valuable information" on financing facilities. Thus, the larger the board size, the more possible are debts.

The CEO duality is positively related to total debt and long-term debt but the correlation between the CEO duality and leverage is insignificant. The result is consistent with Chen et al. (2008) who also discovered that the CEO duality is insignificantly related to firm performance.

The Big-four auditor is barely negatively correlated with short-term debt; this is not consistent with our expectation. However, Oxelheim (2006) also pointed out that the governance disclosure is negatively related to debt ratio. According to Qi et al. (2013), the firms employing high quality auditors could improve the reliability of their financial report in so doing, decrease the information asymmetry, whence leading toward cutting down debt cost.

In terms of the firm factors, the company size is negatively linked with long-term debt. According to Chen (2004), due to their reputation, large companies more easily obtain equity financing support from the secondary market. Moreover, the debt holder does not have the legal safeguard of their debt—because the legal safeguard is undefined in China and the bankruptcy cost is not high. Thus, the debt holders do not likely much appreciate long-term debts of the firm. It might be one of the main reasons why long-term debt is negatively related to the firm size.

There is a positive correlation between the tangibility and total debt and long-term debt, but a negative correlation between the tangibility and short-term debt. The result is the same as that found by Ortiz-Molina and Penas (2008), because the fixed assets are not appropriate for short-term loan guarantee (cited in Dasilas and Papasyriopoulos 2015). Amidu (2007) also finds this negative relationship between short-term debt and tangible assets, in Ghana.

There is an inverse correlation between profitability and total debt. This is not consistent with our expectation. Notice that the result supports the pecking order theory. According to Chen (2004), the reason might stem in the limited banking financial resource, uncompleted bond markets, and a lack of legal protection. Moreover, the equity is more attractive than debt because capital gain is massive and debts are binding.

The non-debt tax shields are negatively correlated to total debt and long-term debt, but positively linked with short-term debt. This means that a company with a larger NDTS is likely to adopt more short-term debt and less long-term debt strategies.

Finally, credit rating is positively correlated with total debt and long-term debt, but is negatively related to short-term debt, indicating that if a company has a high credit rating, the company would hold more long-term debt but less short-term debt. According to Diamond (1991), the company with a low credit rating has no choice, but to adopt the short-term debt strategy. Notice that his theory recognizes two types of short-term borrowers: one is the firm with low credit rating—which has no choice, the other is the firm with high credit rating which adopts its debts to time of borrowing in order to make good use of information arrivals. Borrowers who have ratings "in between" depend more heavily on long-term debts.

*5.2. Sensitivity Tests*

In order to test the results' robustness, some independent factors are to be replaced and thereafter one reruns a regression analysis. For example, profitability can be replaced by the return on equity (ROE), which is the ratio of retained profits to equities. Thus, we impose $\beta_7 = 0$ in the model and take $\beta_{12}$ as finite.

Three dummy factors are used for three credit groups: CR4 is the dummy representing the value of 1 for companies which rate at AAA and 0 otherwise. CR3 is the dummy representing the value of 1 for companies which rate at AA+, AA, AA− and 0 otherwise. CR2 is the dummy representing the value of 1 for companies which rate at A+, A, A− and 0 otherwise.

The results are shown in Table 4, for the regression outputs with ROE and CR terms but without PROFIT term, in Table 5, for the regression outputs with PROFIT term and CR4, CR3, CR2 but without ROE term, and in Table 6, for the regression outputs with ROE term and CR4, CR3, CR2 but without PROFIT term, all for 2010 to 2015. For each regression coefficient, the statistical significance is given, as well as the appropriate *F* test result.

The regression results between company variables and corporate governance factors show similar signs and relationships as those displayed in Table 3. However, the three credit rating groups display different signs. CR4 is positively related to total debt, but both CR3 and CR2 are negatively correlated to total debt. All groups are negatively correlated with short-term debt. Only CR3 presents a positive relationship with long-term debt ratio. Therefore, an important conclusion arises: it can be concluded that there is a different significance between the high-rated credit rating and the low-rated credit rating for a firm capital structure. According to Becker and Milbourn (2008), the "let's go shopping" for a better rating by some more accommodating auditor might be a cause of the weak linear relations between DR and credit rating dummy seen in Table 6.

**Table 4.** Regression outputs with ROE and CR terms but without PROFIT term for 2010 to 2015.

| Variable | Expected Sign | DR | LDR | SDR |
|---|---|---|---|---|
| BOARD | − | 0.01 *** | −0.01 *** | −0.02 *** |
|  |  | (3.40) | (−3.16) | (−7.17) |
| OUTSIDERS | − | 0.35 *** | 0.20 *** | −0.05 |
|  |  | (3.94) | (3.26) | (−0.48) |
| DUALITY | ± | 0.01 | −0.00 | 0.01 |
|  |  | (0.95) | (−0.37) | (0.94) |
| AUDITOR | + | −0.02 | 0.02 | −0.04 ** |
|  |  | (−0.93) | (1.50) | (−2.19) |
| SIZE | + | 0.05 *** | −0.01 *** | 0.06 *** |
|  |  | (11.09) | (−3.69) | (11.57) |
| TANGIBILITY | + | 0.14 *** | 0.38 *** | −0.17 *** |
|  |  | (3.57) | (14.00) | (−4.04) |
| ROE | + | −0.00 *** | −0.00 | −0.00 *** |
|  |  | (−3.68) | (−1.61) | (−3.88) |
| GROWTH | − | −0.03 *** | −0.02 *** | −0.01 * |
|  |  | (−9.65) | (−7.66) | (−1.91) |
| NDTS | ± | −3.84 *** | −3.49 *** | 1.36 ** |
|  |  | (−7.26) | (−9.58) | (2.37) |
| LOGAGE | − | 0.05 *** | 0.02** | 0.01 |
|  |  | (3.53) | (2.17) | (0.42) |
| CR |  | 0.01 * | 0.02 *** | −0.07 *** |
|  |  | (−1.27) | (3.79) | (−6.69) |
| $R^2$ |  | 0.415 | 0.273 | 0.201 |
| *F* Test | $F(16,982) =$ | 34.25 | 22.91 | 15.74 |
| Prob > *F* |  | 00.000 | 0.000 | 0.000 |

Prob > *F* represents the significance of the Joint test; all such test results are 0.000 < 0.01, illustrating that the significance results are over 1% level. *T* test is in parentheses. The *, ** and *** represent statistical significance at the 0.1, 0.05 and 0.01 level, respectively.

**Table 5.** Regression outputs with PROFIT term and CR4, CR3, CR2 but without ROE term for 2010 to 2015.

| Variable | Expected Sign | DR | LDR | SDR |
|---|---|---|---|---|
| BOARD | − | 0.00 * | −0.01 *** | −0.02 *** |
| | | (1.83) | (−3.37) | (−6.95) |
| OUTSIDERS | − | 0.35 *** | 0.22 *** | 0.02 |
| | | (4.13) | (3.54) | (0.22) |
| DUALITY | ± | 0.01 | 0.00 | 0.02 |
| | | (0.80) | (0.00) | (1.20) |
| AUDITOR | + | −0.01 | 0.03 ** | −0.01 |
| | | (−0.84) | (2.26) | (−0.44) |
| SIZE | + | 0.05 *** | −0.01 *** | 0.06 *** |
| | | (11.94) | (−3.49) | (12.50) |
| TANGIBILITY | + | 0.16 *** | 0.38 *** | −0.18 *** |
| | | (4.26) | (14.13) | (−4.30) |
| PROFIT | + | −1.01 *** | −0.16 ** | −0.45 *** |
| | | (−10.45) | (−2.27) | (−4.11) |
| GROWTH | − | −0.02 *** | −0.02 *** | −0.00 |
| | | (−7.84) | (−7.19) | (−1.30) |
| NDTS | ± | −3.76 *** | −3.52 *** | 1.27 ** |
| | | (−7.50) | (−9.77) | (2.26) |
| LOGAGE | − | 0.04 *** | 0.02 ** | 0.00 |
| | | (3.19) | (2.30) | (0.29) |
| CR4 | | 0.05 * | 0.03 | −0.26 *** |
| | | (−1.61) | (1.55) | (−7.85) |
| CR3 | | −0.04 * | 0.04 ** | −0.08 *** |
| | | (−1.66) | (2.20) | (−3.11) |
| CR2 | | −0.05 ** | −0.02 | −0.06 ** |
| | | (−2.03) | (−1.05) | (−2.00) |
| $R^2$ | | 0.415 | 0.273 | 0.201 |
| *F* Test | $F(18,980) =$ | 39.02 | 21.61 | 16.76 |
| Prob > *F* | | 0.000 | 0.000 | 0.000 |

Prob > *F* represents the significance of the Joint test, all such test results are 0.000 < 0.01, illustrating that the significance results are over 1% level. *T* test is in parentheses. The *, ** and *** represent statistical significance at the 0.1, 0.05 and 0.01 level, respectively.

**Table 6.** Regression outputs with ROE term and CR4, CR3, CR2 but without PROFIT term.

| Variable | Expected Sign | DR | LDR | SDR |
|---|---|---|---|---|
| BOARD | − | 0.01 *** | −0.01 *** | −0.02 *** |
| | | (3.12) | (−3.11) | (−6.54) |
| OUTSIDERS | − | 0.34 *** | 0.22 *** | 0.02 |
| | | (3.82) | (3.52) | (0.22) |
| DUALITY | ± | 0.01 | 0.00 | 0.01 |
| | | (1.15) | (0.03) | (1.16) |
| AUDITOR | + | −0.02 | 0.03 ** | −0.01 |
| | | (−1.03) | (2.24) | (−0.41) |
| SIZE | + | 0.05 *** | −0.01 *** | 0.06 *** |
| | | (10.57) | (−3.51) | (12.59) |
| TANGIBILITY | + | 0.14 *** | 0.38 *** | −0.19 *** |
| | | (3.70) | (14.02) | (−4.51) |
| ROE | + | −0.00 *** | −0.00 | −0.00 *** |
| | | (−3.53) | (−1.53) | (−4.15) |
| GROWTH | − | −0.03 *** | −0.02 *** | −0.01 * |
| | | (−9.73) | (−7.65) | (−1.65) |
| NDTS | ± | −3.80 *** | −3.56 *** | 1.08 * |
| | | (−7.18) | (−9.80) | (1.90) |
| LOGAGE | − | 0.05 *** | 0.02 ** | 0.01 |
| | | (3.69) | (2.41) | (0.42) |
| CR4 | + | 0.05 * | 0.03 | −0.25 *** |
| | | (−1.73) | (1.59) | (−7.52) |
| CR3 | − | −0.06 ** | 0.04 ** | −0.08 *** |
| | | (−2.26) | (2.14) | (−2.99) |
| CR2 | − | −0.06 ** | −0.02 | −0.05 * |
| | | (−2.23) | (−1.03) | (−1.83) |
| $R^2$ | | 0.415 | 0.273 | 0.201 |
| *F* Test | $F(18,980) =$ | 30.72 | 21.39 | 16.79 |
| Prob > *F* | | 0.000 | 0.000 | 0.000 |

Prob > *F* represents the significance of the Joint test, all such test results are 0.000 < 0.01, illustrating that the significance results are over 1% level. *T* test is in parentheses. The *, ** and *** represent statistical significance at the 0.1, 0.05 and 0.01 level, respectively.

## 6. Conclusions

The purpose of this study was to examine new relationships, i.e., both corporate governance and credit rating influence on capital structure and leverage conditions of large listed companies on the Shanghai Stock Exchange (SHSE) and Shenzhen Stock Exchange (SZSE) from 2010 to 2015. The 182 company sampling resulted from filtering the data according to credit rating methodology using the assessment of Shanghai Brilliance Credit Rating & Investors Service Company (SBCR) in A-shares. The focus was on firm leverage, considering 3 dependent variables, i.e., Debt Ratio, Long-term Debt Ratio, and Short-term Debt Ratio, in terms of 11 independent variables, deduced, as potentially appropriate from an extensive literature review.

In particular, the present study involves corporate governance factors such as, (i) board size, (ii) outside director, (iii) CEO duality, and (iv) auditor. Furthermore, the study also takes the (v) credit rating into consideration, and whether it would influence capital structure. Moreover, other variables, which concern the possible impact between corporate governance and company-specific profiles, such as (vi) firm size, (vii) asset tangibility, (viii) profitability (measured by the return on equity ratio), (ix) growth opportunity, (x) non-debt tax shields, and (xi) firm age are considered.

This study has tested instrument validity by using the Sargan test of over-identifying limitations in each regression. The results of the relationship between board size, firm size, tangibility and growth opportunity factors and debt match our hypothesis or expectations. On the other hand, a CEO-Board chairship duality is insignificantly related to leverage. The non-debt tax shield is significantly correlated with leverage.

There are also some other variable correlation values that reject our expectations. For instance, both the outside director and firm age are positively linked with leverage; but both the Big-Four auditor and profitability are negatively correlated to leverage. The relationship between credit rating and leverage partly match the assumption, but the relationship is weakly significant.

In conclusion, one may fairly suggest that the relationship between credit rating and capital structure of firms is not yet definitively sorted out; the matter should still be further considered through deeper research, in view of, for example, proposing a more refined model suitable, for emphasizing China's credit rating system impact. This seems an interesting study leading to useful prognosis as well as for quantifying political regulations.

In a further study, one could include the crisis period during 2008 to 2010, comparing the data across the crisis, in order to observe similarities, or not, and thus somewhat universal features. In addition, one could investigate the same set of variables for the small-to-medium enterprises (SMEs) allowing comparison of the influence of different sizes on capital structure.

Another point stems from Kou et al. (2015) who argue that credit rating, in China, does not provide valuable information that strongly influences the cost of capital. However, their samples primarily come from State-owned credit rating agencies and from foreign credit rating agencies, thus likely provide different estimates than those used by the private Chinese credit agency used in this paper. Therefore, choosing the different nature credit rating agencies to see whether they influence the rating results or/and whether there is shopping for better credit rating would be a challenging research development. However, the matter is not easy since one asks to have ratings and justifications on the same company by different auditors—data which is not easily made available.

Nevertheless, at this level, it can be claimed that the study provides some useful information for the company directors and the governments which want to adopt corporate governance as the structure system, or want to operate and control a firm in order to achieve a long-term equilibrium between equities and debts, and therefore reduce the number of bankruptcies. To reach such equilibrium could indeed minimize the results of excessive debts, which put firms at risk. Furthermore, this study, which also wishes to emphasize the significance of credit rating, seems to have reached its goal, to make aware that credit rating is high premium: the company, credit agency and government should face the perspective.